\documentclass[10pt,a4paper,online]{article}

\usepackage{amsmath}
\usepackage{amssymb}
\usepackage{graphicx}
\usepackage[labelfont=bf]{caption}

\usepackage{todonotes}
\usepackage{booktabs}
\usepackage{subcaption}

\usepackage{authblk}

\renewcommand{\vec}[1]{\boldsymbol{#1}}

\usepackage{geometry}
\title{An asymptotic hyperbolic-elliptic model for flexural-seismic metasurfaces}

\author[1*]{P. T. Wootton}
\author[1]{J. Kaplunov}
\author[2]{D. J. Colquitt}

\affil[1]{\textit{School of Computing and Mathematics,
		Keele University, Keele, ST5 5BG, UK}}
\affil[2]{\textit{Department of Mathematical Sciences,
		University of Liverpool,
		Liverpool, L69 7ZL, UK}}
\affil[*]{Corresponding Author: p.t.wootton@keele.ac.uk}

\date{}

\begin{document}
	
\maketitle
\begin{abstract}
We consider a periodic array of resonators, formed from Euler-Bernoulli beams, attached to the surface of an elastic half-space. Earlier studies of such systems have concentrated on compressional resonators. In this paper we consider the effect of the flexural motion of the resonators, adapting a recently established asymptotic methodology that leads to an explicit scalar hyperbolic equation governing the propagation of Rayleigh-like waves. Compared with classical approaches, the asymptotic model yields a significantly simpler dispersion relation, with closed form solutions, shown to be accurate for surface wave-speeds close to that of the Rayleigh wave. Special attention is devoted to the effect of various junction conditions joining the beams to the elastic half-space which arise from considering flexural motion and are not present for the case of purely compressional resonators. Such effects are shown to provide significant and interesting features and, in particular, the choice of junction conditions dramatically changes the distribution and sizes of stop bands. Given that flexural vibrations in thin beams are excited more readily than compressional modes and the ability to model elastic surface waves using the scalar wave equation (i.e. waves on a membrane), the paper provides new pathways toward novel experimental set-ups for elastic metasurfaces.
\end{abstract}

\section{Introduction}
In recent years there has been an explosion of interest, both scholarly and popular, in the use of so-called \emph{locally resonant metamaterials} in the control of wave propagation. The concept has its origins in optics~\cite{shelby2001experimental}, but has since found application across a broad range of fields~\cite{kadic2013metamaterials}, including acoustics~\cite{liu2000locally,li2004double,movchan2004split,guenneau2007acoustic}, elasticity~\cite{zhou2012elastic,milton2007modifications,achaoui2011experimental}, flexural waves in thin plates~\cite{colombi2014sub,williams2015theory,xiao2012flexural}, and seismic waves~\cite{brule2014experiments,colombi2016forests}.
Despite the term \emph{metamaterial} first appearing at the turn of the millennium~\cite{Walser2001}, the development of this relatively new field continues apace.
The essential concept is that, using arrays of local resonators, metamaterials can be designed to have properties that do not occur in materials found in nature; the canonical example is negative refraction~\cite{veselago1968electrodynamics}.

At their inception, metamaterials were designed for the control of waves in bulk media.
This idea can be extended to surface waves, creating so-called \emph{metasurfaces}~\cite{maradudin2011structured}; these meta surfaces have the substantial advantage over metamaterials in that they are relatively easy to manufacture~\cite{chen2016review}.
The majority of work on metasurfaces has focused on electromagnetism~\cite{chen2016review} and, for example, spoof-surface-plasmons~\cite{pendry2004mimicking}.
At present, however, the use of metasurfaces has begun to gain traction in other fields, such as acoustics~\cite{ni2014acoustic,zhou2016precise} and, very recently, elastodynamics~\cite{colombi2016seismic,colquitt2017seismic} and fluid-solid interactions~\cite{skelton2018multi}.
In comparison to, say, electromagnetism and acoustics, the treatment of metasurfaces in the framework of elastodynamics is far more challenging and offers a greater array of unique phenomena.

Despite the singular challenges associated with mechanical metamaterials, they have a substantial advantage in that the range of scales over which they can be utilized is large compared with Electromagnetic metamaterials: from the order of millimeters~\cite{buckmann2014elasto} for cloaking of small-scale structures to several meters~\cite{brule2014experiments} for seismic metamaterials.
The second notable feature is the existence of two bulk waves (shear and pressure), which propagate at two different speeds;
this phenomena is unique to elastic waves and has no analogue in electromagnetics or acoustics.
The existence of two distinct classes of wave adds both challenges and novel features.
Elastic media also support surface waves, again travelling at two different speeds, that are localised to the surface of the medium, but do not decay in the directions parallel to the surface.
This is in contrast to electromagnetism where surface plasmons~\cite{ritchie1957plasma} decay in all directions.
The closest analogue of ``true'' surface waves in electromagnetism are so-called spoof-surface-plasmons~\cite{pendry2004mimicking}, which are supported on structured surfaces.

Recently, there have been several analytical~\cite{colombi2016seismic,colquitt2017seismic}, numerical~\cite{colombi2016forests,palermo2016engineered}, and experimental~\cite{colombi2017enhanced} studies on the use of sub-wavelength resonators for the control of surface waves on elastic substrates.
Notable features of such resonant arrays include filtering, focusing, rainbow trapping, and mode-conversion of surface waves to bulk waves~\cite{colombi2017elastic}.

Interestingly, a recent numerical and experimental study~\cite{rupin2015symmetry} has shown that the hybridisation of the two zeroth-order modes in thin plates can be controlled by using symmetric and asymmetric distributions of resonators.
The experimental studies~\cite{colombi2017enhanced,rupin2015symmetry} are incumbent on prior completion of numerical or analytical studies.
The numerical approaches are computationally intensive and require extensive GPU computing facilities~\cite{colombi2016forests}, while complete analytical studies present their own significant challenges~\cite{colquitt2017seismic}.

In the present paper we develop a novel asymptotic formulation, initially introduced in~\cite{kaplunov2004asymptotic}, to analyse the interaction of surface waves with a resonant array resting on an elastic half-space.
This asymptotic approach is used to analyse the dynamic effects near the surface of semi-infinite elastic bodies and consists of an explicit hyperbolic equation governing the Rayleigh wave propagation as well as a pseudo-static elliptic equation governing decay into the interior.
This approach has proven a powerful tool for a variety of near-surface elastodynamic problems, including 3-dimensional moving load problems~\cite{kaplunov2017asymptotic} and the effect of adding pre-stress to the elastic half-space~\cite{khajiyeva2018hyperbolic}.
For more general works, the reader is referred ~\cite{kaplunov2006explicit,ege2017surface,nobili2018explicit}.
Recently this hyperbolic-elliptic model was applied to a resonant array of rods attached to a half-space ~\cite{ege2018approximate}; these rods were assumed to undergo only longitudinal vibrations as was the case in previous analytical~\cite{colquitt2017seismic} and numerical~\cite{colombi2016seismic} studies.
However recent experimental data obtained for a dense forest of pine trees, acting as subwavelength resonators, suggests that the flexural resonances as well as the junction conditions~\cite{roux2018toward} may play an important role; yet, little attention has thus far been given to the junction condition joining the resonators to the substrate.

Here we extend this approach to include the effects of flexural motion in the thin rods and investigate the effect of these flexural resonances as well as the junction conditions coupling the resonators to the
elastic half-space. In order to better see the effect of the flexural motion, and because in the frequency ranges considered flexural waves dominate, the longitudinal motion  of the resonators is neglected.
These new interactions bring new features, including the ability to control the width of the deep sub-wavelength stop bands.
This is an important feature as, usually, sub-wavelength resonant arrays are only capable of producing very narrow band gaps~\cite{zhou2012elastic} and, as a result, it is challenging to make use of these band gaps in the design of filters and other wave control devices. While a non-dimensionalisation of this problem would be useful to consider, this manuscript is following from an existing body of work with strong practical motivations; in the literature of such problems it is more typical to see the results presented in dimensional form.
	
As is usual for the treatment of such low amplitude bending waves, the effect of the weight of the beams on vertical force into the half-space and the bending motion of the resonators is considered to be negligible compared to the bending forces.

The present paper is be organised as follows.
The problem formulation is given in Section \ref{section:Statement of the problem} with the formulation for Rayleigh waves in subsection \ref{subsec:Asymptotic Formulation for Rayleigh Waves}.
The sections which follow then each detail the treatment of a different junction condition between the resonators and the half-space.
Section \ref{section:Simply Supported Beams} considers the resonators as simply supported, with the resonators able to freely rotate but conserve displacements and horizontal stresses.
In Section \ref{section:Beams on a Rail} the resonators are considered as ideally attached at the base to rails which can move freely but matches the bending moment and the gradient at the surface.
For the junction considered in Section \ref{section:Fully Matched Beams} we consider the resonators as ideally attached to the half-space and so all variables between the resonator and half-space are fully matched.
In each section, the solution obtained from the asymptotic formulation is compared with an implicit evaluation of the full unimodal solution as used in the previous treatment of similar systems \cite{colquitt2017seismic}.

\section{Statement of the problem}
\label{section:Statement of the problem}

\begin{figure}
\centering
\includegraphics[width=0.8\linewidth]{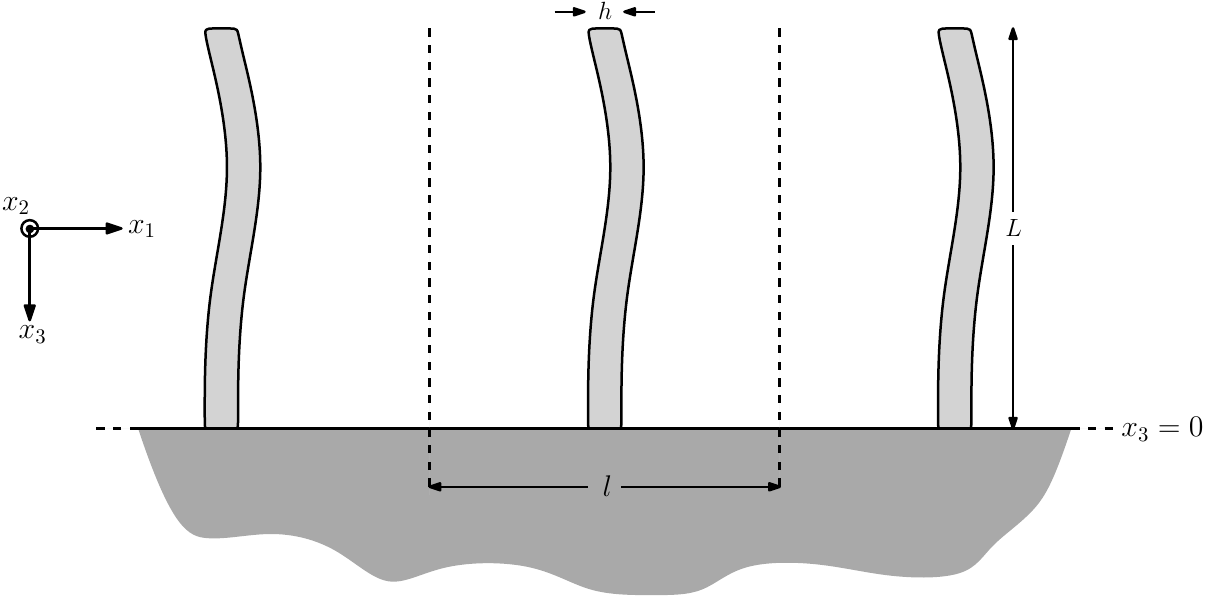}
\caption{A schematic representation of the system under consideration: a periodic array of flexural beams, with lattice constant $\ell$, height $L$, and thickness $h$, on the surface of an elastic half-space.}
\label{fig:schematic}
\end{figure}

Before considering the effects of the various interface conditions, we begin with a statement of the problem.
We consider the propagation of surface waves on an infinite elastic half-space upon which rests an array of thin Euler-Bernoulli beams, as illustrated in Figure~\ref{fig:schematic}.
These surface waves propagate with constant amplitude along the free surface of the half-space and decay, exponentially, into the bulk.
We begin by formulating the three-dimensional problem.

\subsection{An Array of Beams embedded into an Elastic Half-Space}\label{subsec:An Array of Beams embedded into an Elastic Half Plane}

We consider the linearly elastic motion of a three-dimensional half-space ${-\infty<x_1<\infty}$ , ${-\infty<x_2< \infty}$, ${0\leq x_3<\infty}$. For the usual displacement vector $\vec{u} = [u_1,u_2,u_3]^\mathrm{T}$ we consider only motion in the $x_1$ and $x_3$ directions and introduce two displacement potentials $\phi$ and $\psi$ such that
\begin{align}
u_1 = \frac{\partial \phi}{\partial x_1} - \frac{\partial \psi}{\partial x_3},
\qquad
u_3 = \frac{\partial \phi}{\partial x_3} + \frac{\partial \psi}{\partial x_1}.
\label{eq:DisplacementsAsDispPots}
\end{align}
Then for linear elasticity, where $\rho$ is the volume mass density and where $\lambda$ and $\mu$ are Lam\'{e}'s first and second parameters respectively, the equations of motion can be expressed as
\begin{align}
\frac{\partial^2 \phi}{\partial x_1^2} + \frac{\partial^2 \phi}{\partial x_3^2} - \frac{1}{c_1^2} \frac{\partial^2 \phi}{\partial t^2} = 0,
\qquad
\frac{\partial^2 \psi}{\partial x_1^2} + \frac{\partial^2 \psi}{\partial x_3^2} - \frac{1}{c_2^2} \frac{\partial^2 \psi}{\partial t^2} = 0,
\label{eq:LinearElasticityGovEquations}
\end{align}
where $t$ is time and $c_1 = \sqrt{(\lambda + 2 \mu) / \rho}$ and $c_2 = \sqrt{\mu/ \rho}$ are the longitudinal and shear wave speed respectively.

The flexural motion of the thin elastic beams that pattern the surface of the elastic half-space and form the resonant array obey the Euler-Bernoulli beam equation
\begin{align}
E_R I_R \frac{\partial^4 \vec{U}}{\partial x_3^4} = - M_R \frac{\partial^2 \vec{U}}{\partial t^2}, \quad i = 1,2,
\label{eq:BeamGovEquation}
\end{align}
where $\vec{U}(t,x_3) = [U_1,U_2]^\mathrm{T}$ is the displacement vector parallel to the surface of the half-space ${x_3=0}$, for simplicity $E_R$ is introduced to represent the resonator stiffness,
\begin{align}
E_R = \frac{E}{1 - \nu^2}
\end{align}
where $E$ and $\nu$ respectively are the Young Modulus and Poisson ratio of the resonators, $M_R$ is the mass per unit length of the resonators, and $I_R$ is the area moment of inertia of the beams, which for resonators with cylindrical cross-section is given by,
\begin{align}
I_R = \frac{\pi h^4}{64},
\end{align}
where $h$ is the diameter of a beam.
The ends of the beams at $x_3 = -L$ are assumed to be free such that the moments and transverse forces vanish
\begin{equation}
\frac{\partial^3 \vec{U}}{\partial x_3^3} = \vec{0},
\qquad
\frac{\partial^2 \vec{U}}{\partial x_3^2} = \vec{0}.
\label{eq:BeamFreeEndBoundCond}
\end{equation}
For the moment, we will leave the interface condition at the base of each resonator ($x_3=0$), connecting it to the half-space, undefined.
These junction conditions will be defined and studied in subsequent sections.
In general, the interface between one-dimensional and three-dimensional bodies results in a singularly perturbed problem which requires careful consideration of the boundary layer; see, for example,~\cite{kozlov1999asymptotic}.

\label{para:no-compression}
For sufficiently thin resonators the resistance to bending motion is, typically, much smaller than the resistance to compression and therefore, in the linear regime, the flexural and compressional deformations are decoupled at leading order (see, for example,~\cite{kaplunov1998dynamics}).
Moreover~\cite{colquitt2017seismic} suggest that, for the case of resonant arrays on thin plates, the effects of flexural and compressional resonances can be considered independently.
Resonant arrays composed of purely compressional resonators atop a half-space have previously been considered in~\cite{colombi2016seismic,colquitt2017seismic,ege2018approximate}.
However, a detailed analysis of the influence of flexural resonances on the propagation of surface waves is lacking.
Therefore, in the present paper, we focus on the flexural motion of the resonators and consider pure bending only such that vertical motion on the surface of the half-space does not couple to compressional deformation of the resonators.
It is however emphasised that vertical tractions may arise at the surface of the half-space as a result of rotations in the flexural resonators~\cite{kaplunov1998dynamics} and these are taken into account.

\subsection{Asymptotic Formulation for Rayleigh Waves}
\label{subsec:Asymptotic Formulation for Rayleigh Waves}

Considering only waves propagating in the positive $x_1$ direction leads to a simplification of the governing equations of elasticity and subsequently a simplification to the asymptotic model, for which the full 3-dimensional model is given in~\cite{kaplunov2017asymptotic}.

Using the displacement potentials \eqref{eq:DisplacementsAsDispPots} and governing equations \eqref{eq:LinearElasticityGovEquations} from before, we introduce the usual stress tensor $\boldsymbol{\sigma}$, where in the case of a free surface, stress free junction conditions are imposed yielding,

\begin{align}
\sigma_{3i} = 0, \quad i=1,2,3.
\end{align}
In terms of the displacement potentials the stresses are given by

\begin{align}
\begin{split}
\sigma_{31} &= 2 \mu \phi_{,13} + \mu \psi_{,11} - \mu \psi_{,33},\\
\sigma_{33} &= (\lambda + 2 \mu) \phi_{,33} + \lambda \phi_{,11} + 2 \mu \psi_{,13}.
\end{split}
\label{eq:SurfaceStresses}
\end{align}
where, as usual, a subscript of the variable denotes a derivative.
We introduce the following Ans\"atze corresponding to time-harmonic waves of radian frequency $\omega$
\begin{align}
\phi = A_{\phi}\ e^{i(kx_1 - \omega t)-k\alpha x_3},
\qquad
\psi = A_{\psi}\ e^{i(kx_1 - \omega t)-k\beta x_3},
\label{eq:PhiandPsiHarmonicWaveRepresentation}
\end{align}
where $A_\phi $, $ A_\psi$, $k$ and are constants and $\alpha$ and $\beta$ are given by
\begin{align}
\alpha = \sqrt{1 - \left(\frac{\omega}{c_1 k}\right)^2},
\qquad
\beta = \sqrt{1 - \left(\frac{\omega}{c_2 k}\right)^2}.
\label{eq:AlphaBetaNonRayleigh}
\end{align}
Substitution of these Ans\"atze into the free surface junction conditions yields the condition for the existence of a non-trivial solution~\cite{colquitt2017seismic}
\begin{align}
4 \alpha \beta - (1+ \beta^2)^2 =0,
\label{eq:RayleighDeterminant}
\end{align}
from which an implicit expression for the Rayleigh wave speed, denoted by $c_R$, can be obtained.
The dispersion equation for such waves is linear, $\omega = c_R k$, and the Rayleigh decay constants are subsequently given by
\begin{align}
\alpha = \sqrt{1 - \left(\frac{c_R}{c_1}\right)^2} \triangleq \alpha_R,
\qquad
\beta = \sqrt{1 - \left(\frac{c_R}{c_2}\right)^2} \triangleq \beta_R.
\end{align}

In the parameter regime of interest, where the characteristic wavelength of the travelling wave is sufficiently large compared with both the cross-section and separation of the resonators, then, as in previous treatments of such problems \cite{ege2018approximate, slepyan1967strain}, the surface stress associated with the resonant array may be distributed in the form
\begin{align}
\sigma_{31} = \frac{H}{l^2}, \qquad \sigma_{31} = \frac{V}{l^2}
\label{eq:DistributedLoad}
\end{align}
where $H$ and $V$ are the amplitudes of the horizontal and vertical forces respectively, and $l$ is the separation between the resonators.

Each type of junction condition treated in this paper results from the matching of variables between the resonators and the surface of the half-space at $x_3=0$.
The same approach will be used in the treatment of the moments arising from the resonators.
However, we defer this discussion until the need arises in \S\ref{section:Beams on a Rail}.

We now develop the asymptotic model, initially introduced by Kaplunov et al. for a near-Rayleigh surface waves (see~\cite{kaplunov2017asymptotic} and references therein for further details).
This asymptotic model is a leading order perturbation of the equations of motion \eqref{eq:LinearElasticityGovEquations} around the Rayleigh wave eigensolution above \eqref{eq:RayleighDeterminant}.
A perturbation in the form of a small surface stress is imposed which causes a small deviation from the Rayleigh speed.
This expansion is valid in the regime where
\begin{align}
\epsilon = \left|1 - \frac{\omega}{c_R k}\right| \ll1,
\end{align}
and, in particular, the solution is known to be valid only if $\frac{\omega}{k} \sim c_R$.
From inspection of the surface conditions this is equivalent to when both of the surface stresses are small.
The equations for the bulk then reduce to
\begin{align}
\begin{split}
\frac{\partial^2 \phi}{\partial x_3^2} + \alpha_R^2 \frac{\partial^2 \phi}{\partial x_1^2} &= 0,\\
\frac{\partial^2 \psi}{\partial x_3^2} + \beta_R^2 \frac{\partial^2 \psi}{\partial x_1^2} &= 0,
\end{split}
\end{align}
which then lead to the travelling wave solutions \eqref{eq:PhiandPsiHarmonicWaveRepresentation} becoming
\begin{align}
\phi = A_{\phi}\ e^{i(kx_1 - \omega t)-k\alpha_R x_3},
\qquad
\psi = A_{\psi}\ e^{i(kx_1 - \omega t)-k\beta_R x_3}.
\label{eq:PhiandPsinearRayleighWaveRepresentation}
\end{align}
Depending on the applied surface stress, the model is formulated in terms of either the longitudinal or shear wave potential.
If $\sigma_{33}=0$, then it is convenient to work with the shear wave potential and the junction condition along the surface $x_3=0$ is
\begin{align}
\psi_{,11} - \frac{1}{c_R^2}\psi_{,tt} &= -\frac{1+\beta_R^2}{2\mu B}\sigma_{31},
\label{eq:KaplunovSurfBoundHorizontal}
\end{align}
where
\begin{align}
B= \frac{\beta_R}{\alpha_R}(1- \alpha_R^2) + \frac{\alpha_R}{\beta_R}(1-\beta_R^2) - (1-\beta_R^4),
\end{align}
If, on the other hand, $\sigma_{31}=0$ then the junction condition at $x_3=0$ is written in terms of the pressure potential such that
\begin{align}
\phi_{,11} - \frac{1}{c_R^2}\phi_{,tt} &= \frac{1+\beta_R^2}{2\mu B}\sigma_{33}.
\label{eq:KaplunovSurfBoundVertical}
\end{align}
Moreover, on the surface $z=0$, the displacement potentials are related by
\begin{align}
\psi_{,1} = -\frac{2}{1+\beta_R^2} \phi_{,3},
\label{eq:KaplunovDispPotSurfaceRelation}
\end{align}
allowing for the full solution to be obtained.
Not only is this form of the surface equations much simpler than the full unimodal form, but for a known stress they will each only require the solving of a single one-dimensional hyperbolic problem along the surface with the known Rayleigh wave speed.

It is straight-forward to recover the three-dimensional field from the two-dimensional asymptotic model.
In particular, it is necessary to introduce a second shear potential $\psi_2$, corresponding to the stress $\sigma_{32}$, and then simply replace any second order derivatives with respect to $x_1$ with the two dimensional Laplace operator, $\Delta$.
For instance, the vertical surface relation \eqref{eq:KaplunovSurfBoundVertical} in the full three-dimensional form becomes
\begin{align}
\Delta \phi - \frac{1}{c_R^2}\phi_{,tt} &= \frac{1+\beta_R^2}{2\mu B}\sigma_{33}.
\end{align}
The final step in recovering the full three-dimensional formulation is to replace the surface potential relation \eqref{eq:KaplunovDispPotSurfaceRelation} by its three-dimensional equivalents
\begin{align}
\phi_{,i} = \frac{2}{1+\beta_R^2}\psi_{i,3} \ , \quad \phi_{,3} = -\frac{1+\beta_R^2}{2}\left(\psi_{1,1} + \psi_{2,2}\right) \ , \quad i=1,2.
\end{align}

The problem under consideration concerns the effect of a resonant array which, as described earlier, can be treated as a distributed stress on a travelling Rayleigh wave; this makes the asymptotic model introduced here an ideal tool to investigate the problem.
For completeness, however, the solution to the full problem (cf. \S\ref{section:Statement of the problem}\ref{subsec:An Array of Beams embedded into an Elastic Half Plane}) will also be computed numerically and used as a comparison to the asymptotic solution as verification.
We shall now proceed to consider the effect of various interface conditions on the propagation of surface waves over the half-space.

\section{Simply Supported Beams}
\label{section:Simply Supported Beams}

\begin{figure}[h]
	\centering
\begin{subfigure}[t]{0.24\textwidth}
	\centering
	\includegraphics[width=\textwidth]{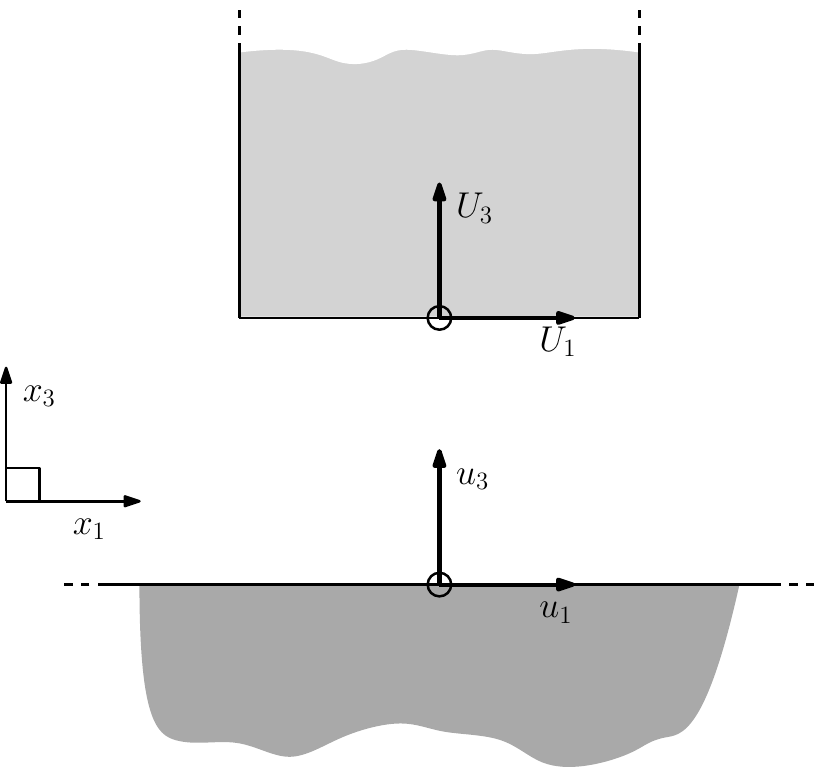}
	\caption{Displacements}
\end{subfigure}
\begin{subfigure}[t]{0.24\textwidth}
	\centering
	\includegraphics[width=\textwidth]{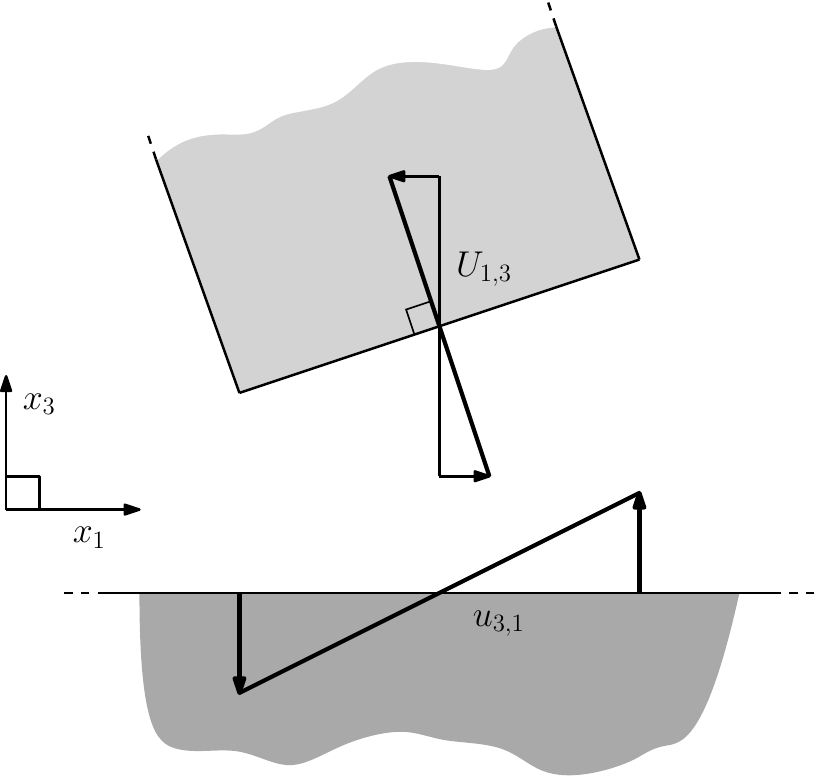}
	\caption{Gradients}
\end{subfigure}
\begin{subfigure}[t]{0.24\textwidth}
	\centering
	\includegraphics[width=\textwidth]{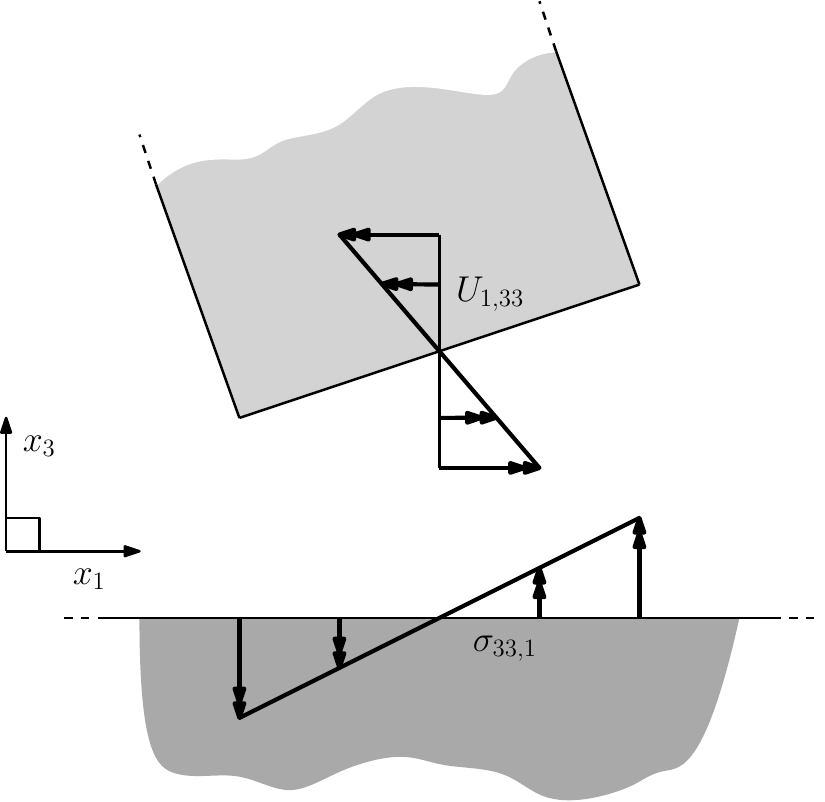}
	\caption{Bending moments}
\end{subfigure}
\begin{subfigure}[t]{0.24\textwidth}
	\centering
	\includegraphics[width=\textwidth]{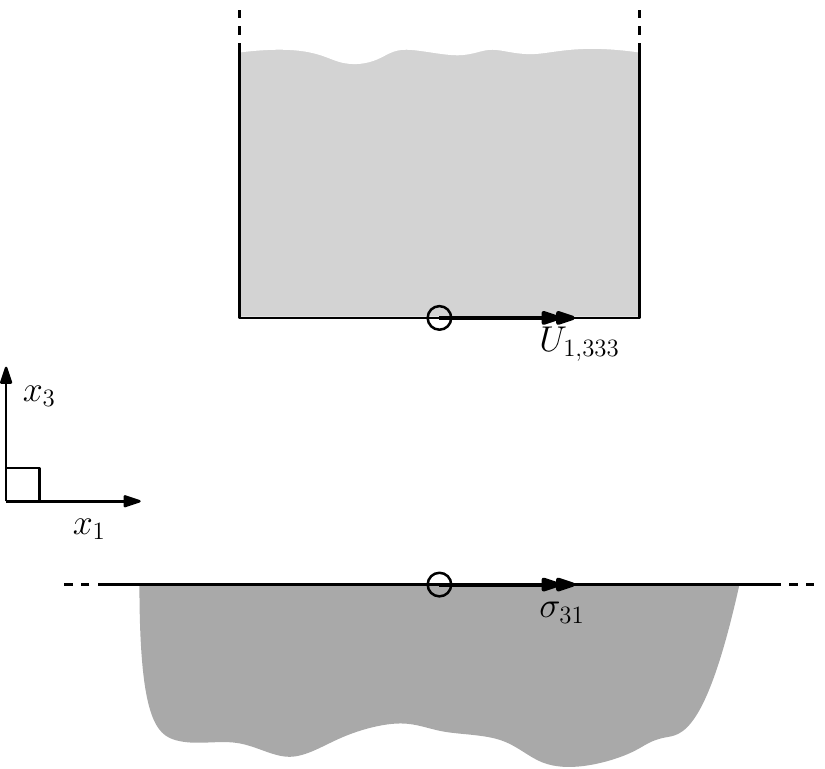}
	\caption{Horizontal stress}
\end{subfigure}

\caption{A schematic representation of the system junction conditions, showing the end of the beam above the edge of the half-space. The relevant half-space quantities and corresponding beam quantities are indicated with a single arrow indicating a dimension of length and a double arrow indicating a dimension of stress.}
\label{fig:boundschematic}
\end{figure}

We begin by considering the resonators as simply supported at the surface of the half-space.
In this system, there is no need to conserve bending angle and there is no overall bending moment at the coupling.
Instead only the horizontal displacement and the transverse force require matching, as illustrated in Figure \ref{fig:boundschematic}.
We emphasise that, as discussed in \S\ref{para:no-compression} on p.\pageref{para:no-compression}, we consider the influence of the resonators under pure bending only and, since there are no rotations at the junction point, the vertical component of traction does not couple into the beam. Note also that, since the beam rotations are not conserved in the half-plane, any rigid body rotations of the beams are also neglected. The junction conditions at the end of the beam atop the half-space are then
\begin{align}
\begin{split}
U_{1}(t,0) &= u_1 |_{x_3=0},\\
U_{1,33}(t,0) &=  0,\\
\end{split}
\label{eq:SimpSuppJunctionConds}
\end{align}
and the stress at the end of the beam is,
\begin{align}
H(x_1,x_2,t) &= \frac{\pi h^4}{64}E_R\ U_{1,333}(t,0) ,
\end{align}
where $U_1$ is the horizontal displacement of the the beam and $H(x_1,y,t)$ is the horizontal force at the surface of the half-space caused by the resonators.

By solving the beam equation \eqref{eq:BeamGovEquation} with these junction conditions and the free end conditions \eqref{eq:BeamFreeEndBoundCond} in the usual way we obtain the relation for horizontal force,
\begin{align}
H(x_1,x_2,t) = E_R \frac{\pi h^4}{64} K^3 \frac{\cos(KL)\cosh(KL) - 1}{\cosh(KL)\sin(KL) - \cos(KL)\sinh(KL)}u_1|_{x_3=0},
\label{eq:SimpSuppH}
\end{align}
where
\begin{align}
K = \left(\frac{64 M_R}{E_R \pi h^4}\right)^\frac{1}{4} \sqrt{\omega}.
\end{align}
This force is applied as a point force in the regular array.
If the size of the array and wavelength of the wave are assumed to be large in comparison with the distance between the point forces then we can homogenise these point forces into a continuous surface stress using the distribution \eqref{eq:DistributedLoad}.
This stress is then given by,
\begin{align}
\sigma_{31} &= \frac{H(x_1,x_2,t)}{l^2},
\end{align}
where $l$ is the distance between two adjacent resonators.

We first consider the full unimodal formulation for the system.
From~\eqref{eq:SurfaceStresses}, the surface stresses can be expressed in terms of displacement potentials such that
\begin{align}
- 2i \ \mu k^2 \alpha \phi - \mu k^2 (1+\beta^2)\psi &= \frac{H(x_1,x_2,t)}{l^2},\\
\left( (\lambda + 2 \mu) k^2 \alpha^2 - \lambda k^2\right) \phi - 2i\mu k^2 \beta \psi &= 0.
\end{align}
From \eqref{eq:DisplacementsAsDispPots}, we have that $U_1(0) = k (i \phi + \beta \psi)$ and, by substituting from \eqref{eq:SimpSuppH}, we obtain the full unimodal dispersion equation
\begin{multline}
\left((\lambda + 2 \mu) k^2 \alpha^2 - \lambda k^2\right)\left(- \mu k^2 \left(1+ \beta^2\right) - k K^3 \hat{H} \beta \right) = \left(2 \ \mu k^2 \alpha + k K^3 \hat{H}\right)\left(-2 \mu k^2 \beta \right),
\label{eq:SimpSuppExactDispRel}
\end{multline}
where $\hat{H} = \sigma_{31}K^{-3}/U_{1}(0)$.
It is clear that in this form this expression will be difficult to manipulate.
If the asymptotic model produces a close approximation which can be used to investigate the behaviour of the dispersion relation, then this will enable us to understand this system on a more fundamental level than the full unimodal solution can.

For the asymptotic model, we use equation~\eqref{eq:KaplunovSurfBoundHorizontal} along the surface,
\begin{align}
\psi_{,11} - \frac{1}{c_R^2}\psi_{,tt} &= -\frac{1+\beta_R^2}{2\mu B l^2}H(x_1,x_2,t),
\end{align}
which, when combined with \eqref{eq:DisplacementsAsDispPots} and \eqref{eq:KaplunovDispPotSurfaceRelation}, yields the asymptotic dispersion relation
\begin{align}
\left( -k^2 + \frac{\omega^2}{c_R^2}\right) \psi = k K^3 \hat{H} \ \beta_R \frac{1-\beta_R^2}{2\mu B} \  \psi.
\end{align}
The asymptotic dispersion relation can therefore be given explicitly, in terms of the roots of the quadratic equation,
\begin{align}
k^2 +k K^3  \hat{H} \ \beta_R \frac{1-\beta_R^2}{2\mu B} \ - \frac{\omega^2}{c_R^2} = 0.
\label{eq:SimpSuppAsymptoticDispRel}
\end{align}
This asymptotic solution can also be obtained directly from the full dispersion relation obtained earlier by expanding around the Rayleigh solution.
In particular, the Rayleigh condition~\eqref{eq:RayleighDeterminant} corresponds to the roots of the function
\begin{align}
R(r) = (2-r)^2 - 4 \sqrt{1-r}\sqrt{1-\frac{c_2^2}{c_1^2}r} = (1+\beta^2)^2 - 4 \alpha \beta,
\label{eq:RayleighFunction}
\end{align}
where $r=\omega^2 c_2^{-2}k^{-2}$.
For $0<r\ll1$, the leading order term of~\eqref{eq:RayleighFunction} is
\begin{align}
R(r) \approx \left(\frac{\omega^2}{c_2^2 k^2}-\frac{c_R^2}{c_2^2}\right) R'\left(\frac{c_R^2}{c_2^2}\right) = -2 \left(1-\frac{\omega^2}{c_R^2 k^2}\right) B.
\end{align}
Therefore, in the case of simply supported beams and small $r$, the full unimodal dispersion relation, \eqref{eq:SimpSuppExactDispRel} can be expanded in the form
\begin{equation}
\frac{K^3 \hat{H}}{\mu k} \beta (1-\beta^2) = \left(2 - \frac{\lambda + 2 \mu}{\mu}(1-\alpha^2)\right)(1+\beta^2) - 4 \alpha \beta = R(r),
\end{equation}
whence
\begin{align}
k^2 - \frac{\omega^2}{c_R^2} = -k\frac{K^3 \hat{H}}{2\mu B} \beta (1-\beta^2),
\end{align}
which is identical to the asymptotic dispersion relation \eqref{eq:SimpSuppAsymptoticDispRel} given that $\beta=\beta_R$.

\begin{figure}[h]
\centering
	\includegraphics[width=0.75\linewidth]{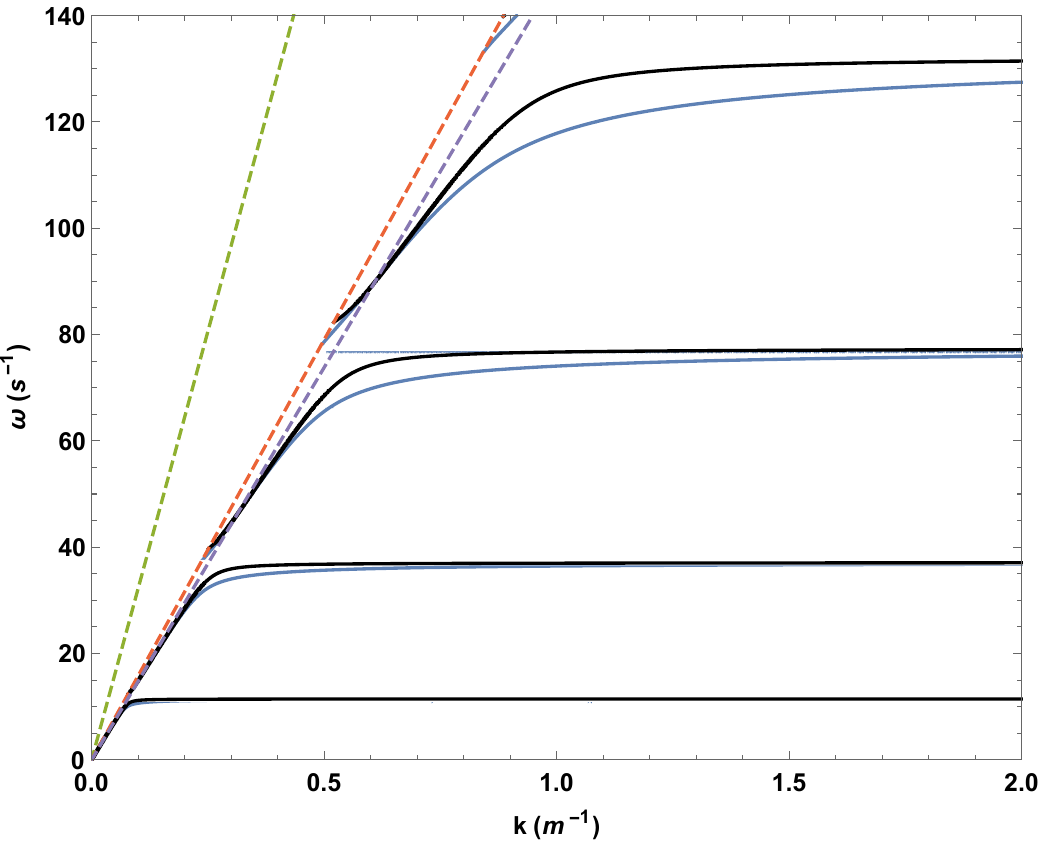}
	\centering
	\caption{The dispersion curves for surface waves on the half-space coated with simply supported beam like resonators, using physical parameters from Table \ref{Tab:BeamsandHalfPlaneParameters}. The solid blue lines show the dispersion curve of the full unimodal solution from \eqref{eq:SimpSuppExactDispRel} and the solid black lines the dispersion curve of the asymptotic solution from \eqref{eq:SimpSuppAsymptoticDispRel}. The dashed green, orange and purple lines correspond to the longitudinal, shear and Rayleigh wave lines respectively.}
	\label{Fig:SimplySupportedBeamResonatorSurfaceWaves}
\end{figure}

It is clear that the asymptotic dispersion relation \eqref{eq:SimpSuppAsymptoticDispRel} is much simpler than the full unimodal equation \eqref{eq:SimpSuppExactDispRel} obtained previously.
Importantly, the asymptotic dispersion relation can be solved explicitly for $k$ as a function of $\omega$ and, therefore, manipulated and interpreted with greater ease.
We compare the asymptotic and full unimodal dispersion relations in Figure \ref{Fig:SimplySupportedBeamResonatorSurfaceWaves} using the parameter values in Table \ref{Tab:BeamsandHalfPlaneParameters}.

\begin{table}[h]
\centering
\caption{The numerical system parameters for the beam resonator and half-space used to produce the dispersion curves shown in Figs. \ref{Fig:SimplySupportedBeamResonatorSurfaceWaves}, \ref{Fig:BeamonaRailResonatorSurfaceWaves}, \ref{Fig:FullyMatchedBeamResonatorSurfaceWaves} and \ref{Fig:FullyMatchedBeamResonatorSurfaceWavesWithDet}.}
\label{Tab:BeamsandHalfPlaneParameters}

\setlength\tabcolsep{7mm}

\begin{tabular}{ccl}
\toprule
Symbol & Definition & Value\\
\midrule
$l$ &Lattice spacing &2 m\\
$\rho$ &Half-space density &13000 kg m$^{-3}$\\
$\mu$ &Half-space shear modulus &325 MPa\\
$\lambda$ &Half-space first Lam\`{e} parameter &702 MPa\\
$L$ &Resonator length &14 m\\
$h$ &Resonator diameter &0.3m\\
$\rho_R$ &Resonator density &450 kg m$^{-3}$\\
$E_R$ &Resonator stiffness & 1.70 GPa\\
\bottomrule
\end{tabular}
\end{table}

The homogenisation scheme from \eqref{eq:DistributedLoad} assumes that the wavelength is on a larger scale than the separation of the resonators, or equivalently $k l \ll 2\pi$. Subtituting $l$ from the the system parameters in Table \ref{Tab:BeamsandHalfPlaneParameters}, this condition is $k \ll \pi \mathrm{m}^{-1}$ for each of the illustrative figures, Figures \ref{Fig:SimplySupportedBeamResonatorSurfaceWaves}, \ref{Fig:BeamonaRailResonatorSurfaceWaves}, \ref{Fig:FullyMatchedBeamResonatorSurfaceWaves} and \ref{Fig:FullyMatchedBeamResonatorSurfaceWavesWithDet}.
In addition, while the asymptotic method gives a solution which is valid for all $k$, we note that in the full unimodal system $\beta$ will become purely imaginary when $\omega/k > c_2$ and so no propagating solution can exist in this region of the plot.

Figure~\ref{Fig:SimplySupportedBeamResonatorSurfaceWaves} shows good agreement between the dispersion curves for the asymptotic and full unimodal approaches; particularly near the Rayleigh sound-line and for lower frequencies, as expected.
In addition, the asymptotes created by the flexural resonances of the beams extends the range of validity of the asymptotic approach beyond the regime of small $k$ and small stresses where one would expect the method to remain accurate.
The flexural resonances associated with these asymptotes create fine deep sub-wavelength band gaps which have been shown to have a range of physical applications (see, for example,~\cite{colquitt2017seismic,colombi2016seismic,colombi2017enhanced,colombi2014sub}).

\section{Beams on a Rail}
\label{section:Beams on a Rail}

Suppose now that the resonators are held at $z=0$ to a rail parallel to the surface of the half-space, which allows the base of the beam to move freely but changes the bending angle of the beam depending on the gradient of the surface.
This allows the beam to impart a bending moment on the half-space but no horizontal forcing.
There is therefore no need to conserve displacements and the horizontal stress must vanish.
The reader is reminded that, as discussed in \S\ref{para:no-compression} on p.\pageref{para:no-compression}, we consider the influence of the resonators under pure bending only.
However, in contrast to the case of simple support beams considered in \S\ref{section:Simply Supported Beams}, the resonators exert a vertical force on the half-space through rotations in the beams.
These rotations can be linked to the gradient of the vertical displacement through the kinematic continuity conditions imposed at the junction points.
Whilst this junction condition may initially appear somewhat counter-intuitive, it can be understood in terms of so-called \emph{gyroscopic hinges}~\cite{nieves2018vibrations}.

To construct the junction conditions for the base of the beam, we must first determine the bending moment $M$ in the elastic half-space caused by the flexural deformation of the beam.
For a beam with cross-sectional area $A$ the bending moment is
\begin{equation}
M = \iint\limits_A x \sigma_{33}dA.
\end{equation}
If the cross-sectional area of the beam is small then the stress can be approximated by the expansion
\begin{equation}
\sigma_{33} \sim \sigma_{33}|_{x=y=0} +x \sigma_{33,1}|_{x=y=0},
\end{equation}
which, on evaluation of the integral, leaves only the second term. This therefore represents the leading order contribution in the calculation of the moments.
Figure \ref{fig:boundschematic} illustrates how this bending moment in the half-space corresponds to the moment at the end of the beam. Furthermore, this figure also shows how, from elementary geometry, the vertical gradient of the half plane is equal to the horizontal gradient of the beam. Together these lead to the junction conditions
\begin{align}
\begin{split}
U_{1,3} (t,0) & = u_{3,1},\\
E_R \frac{\pi h^4}{64} U_{1,33}(t,0) &=  -\frac{\pi h^4}{64} \sigma_{33,1}|_{x_3=0} ,\\
U_{1,333}(t,0) &= 0.\\
\end{split}
\label{eq:BeamRailJunctionConds}
\end{align}
The vertical stress at the junction point is
\begin{align}
V(x_1,x_2,t) &=\frac{\pi h^2}{4}\sigma_{33}|_{x_3=0},
\end{align}
where $V(x_1,x_2,t)$ is the vertical force exerted on the half-space by the beams and $u_{3,1}$ is the gradient of the vertical displacement along the surface.
In particular, 
solving these in the usual way along with the free end conditions \eqref{eq:BeamFreeEndBoundCond} and the beam equation \eqref{eq:BeamGovEquation} gives the vertical force,
\begin{align}
V(x_1,x_2,t) = i \frac{K}{k} \frac{\pi h^2}{4} E_R\ \frac{1-\cos(KL)\cosh(KL)}{\cosh(KL)\sin(KL) + \cos(KL)\sinh(KL)}\,u_{3,1}|_{x_3=0}
\end{align}
We will again start with the full unimodal treatment.
As we did previously with $H(x_1,x_2,t)$, we will use the force distribution \eqref{eq:DistributedLoad} and define $\hat{V}$ such that,
\begin{align}
\frac{V(x_1,x_2,t)}{l^2} &= i \frac{K}{k}\hat{V} u_{3,1}.
\label{eq:Vhat WithStressAndDisplacement}
\end{align}
Expressing $u_{3,1}$ in terms of the potentials~\eqref{eq:DisplacementsAsDispPots} and making use of~\eqref{eq:SurfaceStresses} the surface stresses on the half-space can be written in the form
\begin{align}
- 2i \ \mu k^2 \alpha \phi - \mu k^2 (1+\beta^2)\psi &= 0,\\
\left( (\lambda + 2 \mu) k^2 \alpha^2 - \lambda k^2\right) \phi - 2 i\ \mu k^2 \beta \psi&= - K \hat{V} (-k \alpha \phi + i k \psi),
\end{align}
whence, the full unimodal dispersion equation is obtained in the form
\begin{multline}
\left((\lambda + 2 \mu) k^2 \alpha^2 - \lambda k^2 -k K \hat{V}\alpha\right)\left(- \mu k^2 \left(1+\beta^2\right) \right)
= \left(2 \ \mu k^2 \alpha\right)\left(-2 \mu k^2 \beta + k K \hat{V}\right)
\label{eq:BeamonaRailExactDispRel}
\end{multline}
As might be expected, this dispersion relation is very similar to the one obtained for the simply supported case and is likewise difficult to manipulate and interpret.
Therefore, in order to provide a clearer understanding of the system, we once again turn to the asymptotic approach.

To apply the asymptotic model to this system, where there is only a vertical stress, we will make use of~\eqref{eq:KaplunovSurfBoundVertical} along the surface of the half-space, whence we obtain
\begin{align}
\phi_{,11} - \frac{1}{c_R^2} \phi_{,tt} = \frac{1+ \beta_R^2}{2 \mu B} V(x_1,x_2,t),
\end{align}
or equivalently, using \eqref{eq:KaplunovDispPotSurfaceRelation} and considering time-harmonic waves, in terms of $\psi$
\begin{align}
\psi_{,11} - \frac{1}{c_R^2} \psi_{,tt} = -i\frac{\alpha_R}{\mu B} V(x_1,x_2,t).
\label{eq:KaplunovSurfBoundVerticalwrtPsi}
\end{align}
Combining the surface potential relation \eqref{eq:KaplunovDispPotSurfaceRelation} with $\hat{V}$ from \eqref{eq:Vhat WithStressAndDisplacement} and the surface junction condition~\eqref{eq:KaplunovSurfBoundVerticalwrtPsi}, we obtain the asymptotic dispersion relation
\begin{align}
k^2 - k\ \frac{\alpha_R}{\mu B} K\hat{V} \frac{1-\beta_R^2}{2} - \frac{\omega^2}{c_R^2} =0.
\label{eq:BeamonaRailAsymptoticDispRel}
\end{align}
As before in \S\ref{section:Simply Supported Beams}, the asymptotic dispersion relation can be obtained by expressing the full unimodal dispersion relation \eqref{eq:BeamonaRailExactDispRel} in terms of $R(r)$
\begin{align}
R(r) = -\frac{K \hat{V}}{\mu k} \alpha(1-\beta^2),
\end{align}
and using the leading order expansion of $R(r)$ for $r\ll 1$, that is, for wave speeds close to that of a pure Rayleigh wave, to obtain
\begin{align}
k^2 - \frac{\omega^2}{c_R^2} = k\frac{K \hat{V}}{2\mu B} \alpha (1-\beta^2),
\end{align}
which is consistent with the asymptotic dispersion relation \eqref{eq:BeamonaRailAsymptoticDispRel} if $\alpha = \alpha_R$ and $\beta = \beta_R$, i.e. the decay rates are equal to those of pure Rayleigh waves.

\begin{figure}
	\includegraphics[width=0.75\linewidth]{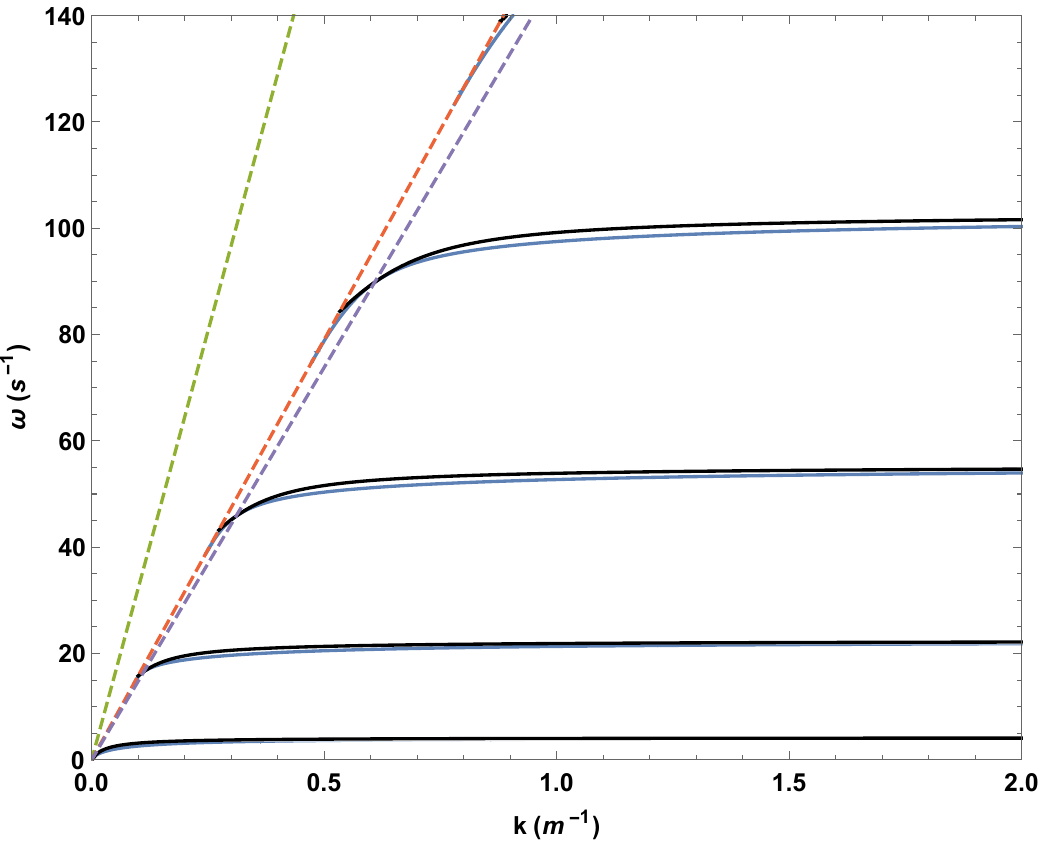}
	\centering
	\caption{The dispersion curves for surface waves on the half-space coated with beam like resonators supported by rails, using physical parameters from Table \ref{Tab:BeamsandHalfPlaneParameters}. The solid blue lines show the dispersion curve of the full unimodal solution from \eqref{eq:BeamonaRailExactDispRel} and the solid black lines the dispersion curve of the asymptotic solution from \eqref{eq:BeamonaRailAsymptoticDispRel}. The dashed green, orange and purple lines correspond to the longitudinal, shear and Rayleigh wave lines respectively.}
	\label{Fig:BeamonaRailResonatorSurfaceWaves}
\end{figure}

As with the simply supported case, the dispersion relation derived from the asymptotic approach is much simpler than the one produced by the complete analysis.
Indeed, the asymptotic dispersion equation is again quadratic and therefore yields an explicit expression for $k$ in terms of $\omega$.
A plot of the solutions of the asymptotic dispersion equation along with those of the full unimodal solutions is shown in Figure~\ref{Fig:BeamonaRailResonatorSurfaceWaves};
the parameter values used are again those detailed in Table~\ref{Tab:BeamsandHalfPlaneParameters}.
Figure~\ref{Fig:BeamonaRailResonatorSurfaceWaves} shows that, once again, there is good agreement between the asymptotic model and the full unimodal solution, particularly near the Rayleigh line.
For all of the modes shown in Figure~\ref{Fig:BeamonaRailResonatorSurfaceWaves}, the asymptotic approach predicts the asymptote, the overall shape of the dispersion relation, and the existence of a large band gap for each mode.
Indeed, in addition to being a much more elegant approach, it is clear that the asymptotic model is capable of reproducing the main features of the full unimodal solution.
Moreover, compared with Figure~\ref{Fig:SimplySupportedBeamResonatorSurfaceWaves}, Figure~\ref{Fig:BeamonaRailResonatorSurfaceWaves} exhibits significantly larger band gaps.

\section{Fully Matched Beams}
\label{section:Fully Matched Beams}
	
We now consider the case where both displacements and rotations (surface gradients), in addition to, horizontal and vertical tractions are coupled at the junction points.
Although, as discussed in \S\ref{para:no-compression} on p.\pageref{para:no-compression} we only consider flexural motion of the resonators and neglect the compressional deformation, the vertical stresses in the half-space couple to the flexural modes in the beams through rotations at the junction point, as in \S\ref{section:Beams on a Rail}; this is fully taken into account here.
Predictably, this complex coupling leads to a more involved solution for both the beam motion and the surface wave dispersion equation. For simplicity, we also assume an ideal contact between the beams and half-space, neglecting any local contact strain effects.

In this case, the equations of motion for the beams~\eqref{eq:BeamGovEquation} together with the free end conditions~\eqref{eq:BeamFreeEndBoundCond} remain unchanged.
The junction conditions at the base of the beam $x_3=0$ are a combination of those previously considered (\ref{eq:SimpSuppJunctionConds}, \ref{eq:BeamRailJunctionConds}) with each of the quantities illustrated in Figure \ref{fig:boundschematic} conserved, giving,
\begin{align}
\begin{split}
U_{1}(t,0) &= u_1|_{x_3=0},\\
U_{1,3}(t,0) & = u_{3,1}|_{x_3=0},\\
E_R \frac{\pi h^4}{64} U_{1,33} |_{x_3=0} &=  -\frac{\pi h^4}{64} \sigma_{33,1}|_{x_3=0} ,\\
\end{split}
\end{align}
corresponding to continuity of displacement, rotation, and the balance of moments respectively.
The associated stresses at the base of the beam are
\begin{align}
\begin{split}
H(x_1,x_2,t) &= \frac{\pi h^4}{64}E_R\ u_{1,333}|_{x_3=0},\\
V(x_1,x_2,t) &= \frac{\pi h^2}{4}\sigma_{33}|_{x_3=0}.
\end{split}
\end{align}
Solving these conditions in the usual way along with (\ref{eq:BeamGovEquation},\ref{eq:BeamFreeEndBoundCond}) gives,
\begin{align}
\begin{split}
H(x_1,x_2,t) &= \frac{\pi h^4}{64} E_R K^2 (\xi u_{3,1} + K \eta u_1) ,\\
V(x_1,x_2,t) &=  i \frac{\pi h^2}{4} E_R \frac{K}{k} (\zeta u_{3,1} + K\xi u_1),
\end{split}
\label{eq:FullyMatchedHandV}
\end{align}
where, for convenience, we have introduced the quantities
\[
\xi = \frac{\sin(KL)\sinh(KL)}{\cos(KL)\cosh(KL) +1},\;
\eta = \frac{\cosh(KL)\sin(KL)+\cos(KL)\sinh(KL)}{\cos(KL)\cosh(KL) +1},
\]
and
\[
\zeta = \frac{\cosh(KL)\sin(KL)-\cos(KL)\sinh(KL)}{\cos(KL)\cosh(KL) +1}.
\]
As in the previous sections, we will begin with the full unimodal treatment and then develop the asymptotic model for comparison.

In contrast to the simply supported and rail conditions discussed in \S\ref{section:Beams on a Rail}~\&~\ref{section:Simply Supported Beams}, we should not expect the full unimodal solution to intersect with the asymptotic solution at the Rayleigh line.
In order for the asymptotic and full unimodal solutions to intersect at the Rayleigh line, the beam must behave as though the coupled end with the half-space is free;
this imposes requirements on the displacement and rotation at the base of the beam that are incompatible with the required displacement and gradient for the half-space.
We emphasise that this feature does not affect the full unimodal solution but does affect the accuracy of the asymptotic solution since the basic assumption underlying the asymptotic model, that stresses tend to zero at the Rayleigh line, is no longer valid.

To formulate the full unimodal problem we begin by writing the surface stresses in terms of potentials \eqref{eq:SurfaceStresses}, and in terms of the distributed load \eqref{eq:DistributedLoad}, to obtain
\begin{align}
\begin{split}
- 2i \ \mu k^2 \alpha \phi - \mu k^2 (1+\beta^2)\psi &= \frac{H(x_1,x_2,t)}{l^2},\\
\left( (\lambda + 2 \mu) k^2 \alpha^2 - \lambda k^2\right) \phi - 2 i\ \mu k^2 \beta \psi &= \frac{V(x_1,x_2,t)}{l^2}.
\end{split}
\label{eq:FullyMatchedSurfaceStresses}
\end{align}
Writing the displacements in terms of their potentials~\eqref{eq:DisplacementsAsDispPots} \&~\eqref{eq:PhiandPsiHarmonicWaveRepresentation} and combining~\eqref{eq:FullyMatchedSurfaceStresses} with~\eqref{eq:FullyMatchedHandV}, we obtain the dispersion equation
\begin{multline}
\left[2\mu k^2 \alpha + \frac{\pi  h^4}{64 l^2}{E_R}
k \left(K^3 \eta -K^2 k \alpha \xi\right)\right] \left[2 \mu k^2 \beta + \frac{\pi  h^2}{4 l^2} E_R \left(K^2 \beta \xi - K k \zeta\right) \right] = \\
\left[\mu k^2 \left(1+\beta^2\right) + \frac{\pi  h^4}{64 l^2} E_R k \left(K^3 \beta \eta - K^2k \xi \right)\right]\\
\times
\left[(\lambda + 2 \mu)k^{2}\alpha^2 - \lambda k^2 + \frac{\pi  h^2}{4 l^2} E_R \left(K^2 \xi - K k \alpha  \zeta\right)\right].
\label{eq:FullyMatchedExactDispRel}
\end{multline}
As one might expect, this dispersion equation is significantly longer and more complicated than those obtained for the previous cases and, correspondingly, more difficult to interpret.
Therefore we will, once again, develop an accurate approximation using our asymptotic mode in an effort to obtain a more transparent dispersion equation.

The asymptotic model of Kaplunov et al.~\cite{kaplunov2004asymptotic} consists of two relations which each employ a single surface stress to produce a wave.
However, in the present problem, the system has two stresses; one of which is normal to, and the other of which is tangential to, the surface.
Therefore, the approach of Kaplunov et al.~\cite{kaplunov2004asymptotic} must be adapted to account for the composition of two stresses into a single wave.
In particular, we will assume that there will be two waves, each generated by one of the two stress-states. Since, for the asymptotic model both waves have a speed which is a small perturbation from the Rayleigh speed, it can also be assumed that these two waves travel at the same speed.
Each of these two waves will have a corresponding pair of displacement potentials, $\psi_1$ and $\phi_1$, and $\psi_2$ and $\phi_2$ such that by linear superposition the overall displacement potentials $\phi$ and $\psi$ are given by a linear combination such that $\phi = \phi_1 + \phi_2$ and $\psi = \psi_1 + \psi_2$.
Each pair of potentials then corresponds to one of the surface relations from (\ref{eq:KaplunovSurfBoundHorizontal}, \ref{eq:KaplunovSurfBoundVerticalwrtPsi}),
\begin{align}
\psi_{1,11} - \frac{1}{c_R^2}\psi_{1,tt} &= -\frac{1+\beta_R^2}{2\mu B l^2}H(x_1,x_2,t),\\
\psi_{2,11} - \frac{1}{c_R^2}\psi_{2,tt} &= -i \frac{\alpha_R}{\mu B l^2}V(x_1,x_2,t).
\end{align}
Hence, assuming time-harmonic waves such that the potentials take the form~\eqref{eq:PhiandPsinearRayleighWaveRepresentation}, we obtain
\begin{align}
\psi_{,11} - \frac{1}{c_R^2}\psi_{,tt} &= -\frac{1+\beta_R^2}{2\mu B}\sigma_{31} -i\frac{\alpha_R}{\mu B}\sigma_{33}.
\label{eq:MultiStressSurfCond}
\end{align}
Combining~\eqref{eq:FullyMatchedHandV} with~\eqref{eq:MultiStressSurfCond}, after writing the displacements in terms of potentials, yields the following dispersion equation
\begin{multline}
0 = k^2\left(1+ \frac{1+\beta_R^2}{2\mu B l^2}\frac{\pi h^4}{64} E_R K^2 \frac{1 - \beta_R^2}{2}\ \xi\right)\\
- k \left( \frac{\alpha_R}{\mu B l^2} \frac{\pi h^2}{4} E_R K \frac{1 - \beta_R^2}{2}\ \zeta - \frac{1+\beta_R^2}{2\mu B l^2}\frac{\pi h^4}{64} E_R K^3 \beta_R \frac{1- \beta_R^2}{1+\beta_R^2}\ \eta \right)\\
-\left(\frac{\omega^2}{c_R^2} + \frac{\alpha_R}{\mu B l^2} \frac{\pi h^2}{4} E_R K^2 \beta_R \frac{1- \beta_R^2}{1+\beta_R^2}\ \xi \right).
\label{eq:FullyMatchedAsymptoticDispRel} 
\end{multline}

\begin{figure}[h!]
	\includegraphics[width=0.75\linewidth]{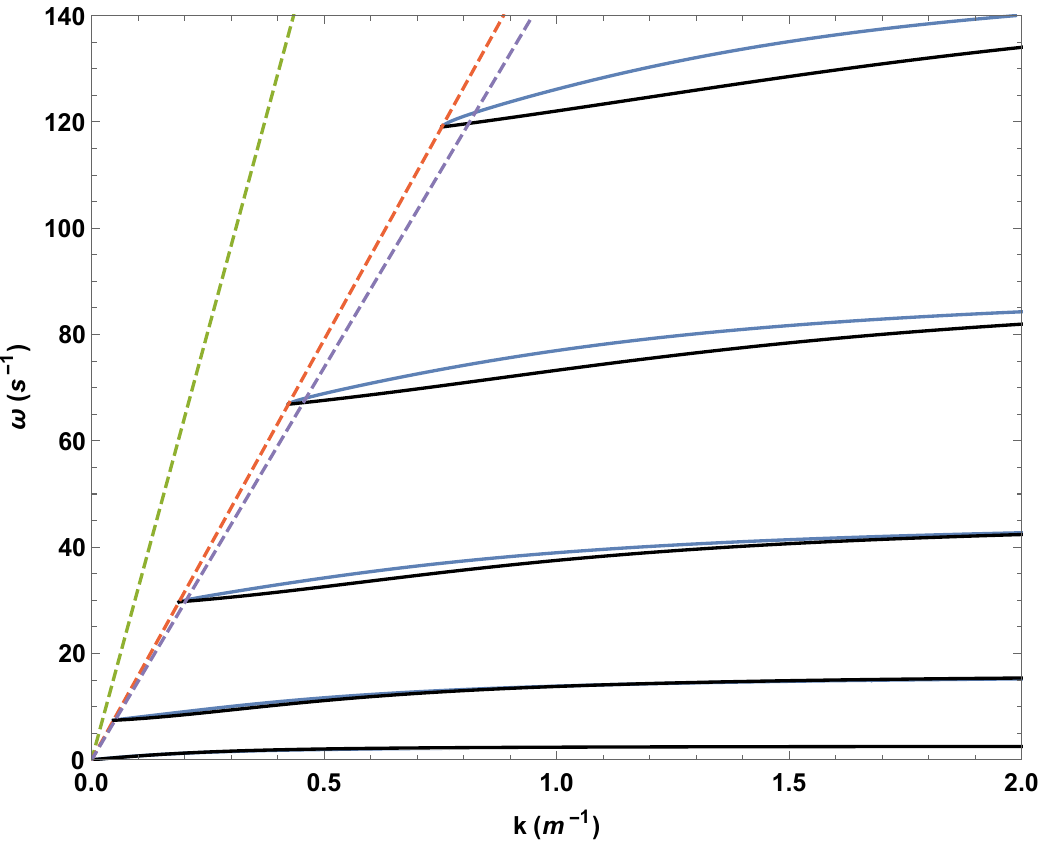}
	\centering
	\caption{The dispersion curves for surface waves on the half-space coated with fully matched beam like resonators, using physical parameters from Table \ref{Tab:BeamsandHalfPlaneParameters}. The solid blue lines show the dispersion curve of the full unimodal solution from \eqref{eq:FullyMatchedExactDispRel} and the solid black lines the dispersion curve of the asymptotic solution from \eqref{eq:FullyMatchedAsymptoticDispRel}. The dashed green, orange and purple lines correspond to the longitudinal, shear and Rayleigh wave lines respectively.}
\label{Fig:FullyMatchedBeamResonatorSurfaceWaves}
\end{figure}

As with the previous cases, we note that the asymptotic dispersion equation~\eqref{eq:FullyMatchedAsymptoticDispRel} is quadratic in $k$ and can therefore be solved explicitly.
Whilst more difficult to interpret than the simply supported asymptotic solution, it is still a substantial improvement on the full unimodal dispersion relation.
Figure \ref{Fig:FullyMatchedBeamResonatorSurfaceWaves} shows the dispersion curves for corresponding to both the asymptotic~\eqref{eq:FullyMatchedAsymptoticDispRel} and full unimodal~\eqref{eq:FullyMatchedExactDispRel} solutions.
We observe that the asymptotic model predicts the main features of the dispersion curves, including prediction of the existence and general shape of the solution branch for each quasi periodic mode and also giving a close approximation of the intersections with the Rayleigh line.
The key feature of the plot, and one present in both the asymptotic and full unimodal results, is the band gaps.
This system is notable for the large range of frequencies which do not have a corresponding real wave number solution and will, therefore, not propagate through the system.
While it is difficult to locate where the band gaps are from the full unimodal dispersion relation, they can be approximated by the intersections between the asymptotic model and the shear wave line.
It is emphasised that, once again, this feature is absent in the previous analyses~\cite{colombi2016seismic,colquitt2017seismic,ege2018approximate} as the flexural resonances of the beams were neglected.

This system also highlights some of the limitations of this model.
In particular, and unlike for the simple supported and rail junctions conditions that involve a single stress, the asymptotics and the full unimodal solution are not guaranteed to intersect at the Rayleigh line.
This is due to the way the stresses are treated in the two methods.
In the asymptotic model there is no interaction between the two stresses and there will be Rayleigh line intersections only at,
\begin{align}
(1+\beta_R^2)\sigma_{31} = 2i\ \alpha_R\ \sigma_{33},
\end{align}
whereas for the full unimodal dispersion relation there is a cross multiplication of the two stress terms.
As discussed earlier, the full unimodal and asymptotic curves will intersect at the Rayleigh line only if the total resultant stress at the junction point is zero.
This, perhaps counter-intuitive, point can be seen by assuming that the two stresses have the form,
\begin{align}
\begin{split}
\sigma_{31} &= H_\phi \phi + H_\psi \psi,\\ 
\sigma_{33} &= V_\phi \phi + V_\psi \psi.
\end{split}
\end{align}
Expanding these stresses as before yields,
\begin{align}
-R(r) =  i \frac{2 \beta}{\mu k^2} H_\phi + \frac{1+\beta^2}{\mu k^2} H_\psi -\frac{1+\beta^2}{\mu k^2} V_\phi + i \frac{2 \alpha}{\mu k^2} V_\psi + \frac{H_\phi V_\psi - V_\phi H_\psi}{\mu^2 k^4}.
\end{align}
If we assume a coincidence at the Rayleigh line then, at this point, we replace $\alpha$ and $\beta$ with $\alpha_R$ and $\beta_R$ and use the surface relation \eqref{eq:KaplunovDispPotSurfaceRelation} to obtain,
\begin{align}
-\left(k^2 - \frac{\omega^2}{c_R^2}\right)\psi = -\frac{1+\beta_R^2}{2 \mu B} \sigma_{31} - i \frac{\alpha_R}{\mu B} \sigma_{33} + \frac{V_\phi H_\psi - H_\phi V_\psi}{2 \mu^2 B k^2} \psi.
\label{eq:MultiStressSurfCondWithDet}
\end{align}
Clearly this equation is equivalent to the multi stress junction condition \eqref{eq:MultiStressSurfCond} but with an added stress interaction term which the previous asymptotic model does not predict.
For this example we can take the surface stresses \eqref{eq:FullyMatchedHandV} and put them into the required form which yields,
\begin{align}
\begin{split}
H_\phi = i \frac{\pi h^4}{64 l^2}E_R K^2 k(K \eta - k\alpha \zeta), \ \qquad &H_\psi = \frac{\pi h^4}{64 l^2}E_R K^2 k (\beta K \eta - k \xi),\\
V_\phi = \frac{\pi h^2}{4 l^2} E_R K(k\alpha\zeta - K \xi), \qquad \qquad &V_\psi = i \frac{\pi h^2}{4 l^2}(\beta K \xi - k \zeta).
\end{split}
\end{align}
Letting $\alpha = \alpha_R$ and $\beta = \beta_R$ and substituting these into the expansion \eqref{eq:MultiStressSurfCondWithDet} leads to the dispersion relation,
\begin{multline}
k^2 - \frac{\omega^2}{c_R^2} = \frac{\alpha_R}{\mu B l^2} \frac{\pi h^2}{4} E_R \frac{K}{k} \left(k^2 \frac{1 - \beta_R^2}{2}\ \zeta + K k \beta_R \frac{1- \beta_R^2}{1+\beta_R^2}\ \xi\right)\\
-\frac{1+\beta_R^2}{2\mu B l^2}\frac{\pi h^4}{64} E_R K^2 \left(k^2 \frac{1 - \beta_R^2}{2}\ \xi + K k \beta_R \frac{1- \beta_R^2}{1+\beta_R^2}\ \eta\right)\\
-\frac{\pi h^6}{256 l^4}\frac{E_R^2 K^3}{2 \mu^2 k}\left((k\alpha_R\zeta - K \xi)(\beta_R K \eta - k \xi) + (\beta_R K \xi - k \zeta)(K \eta - k\alpha_R \zeta)\right),
\label{eq:FullyMatchedAsymptoticDispRelWithDet} 
\end{multline}
which is the same as \eqref{eq:FullyMatchedAsymptoticDispRel} but with an added term from the stress interaction.
Due to cancellation in the stress interaction term this dispersion relation is quadratic on $k$.
However, even for simple stresses, the dispersion equation arising from the aforementioned stress interactions will, in general, be a quartic polynomial.

\begin{figure}[h!]
\centering
\setlength{\fboxrule}{0.5pt}
\begin{tikzpicture}
\node at (0,0){
	\includegraphics[width=0.75\linewidth]{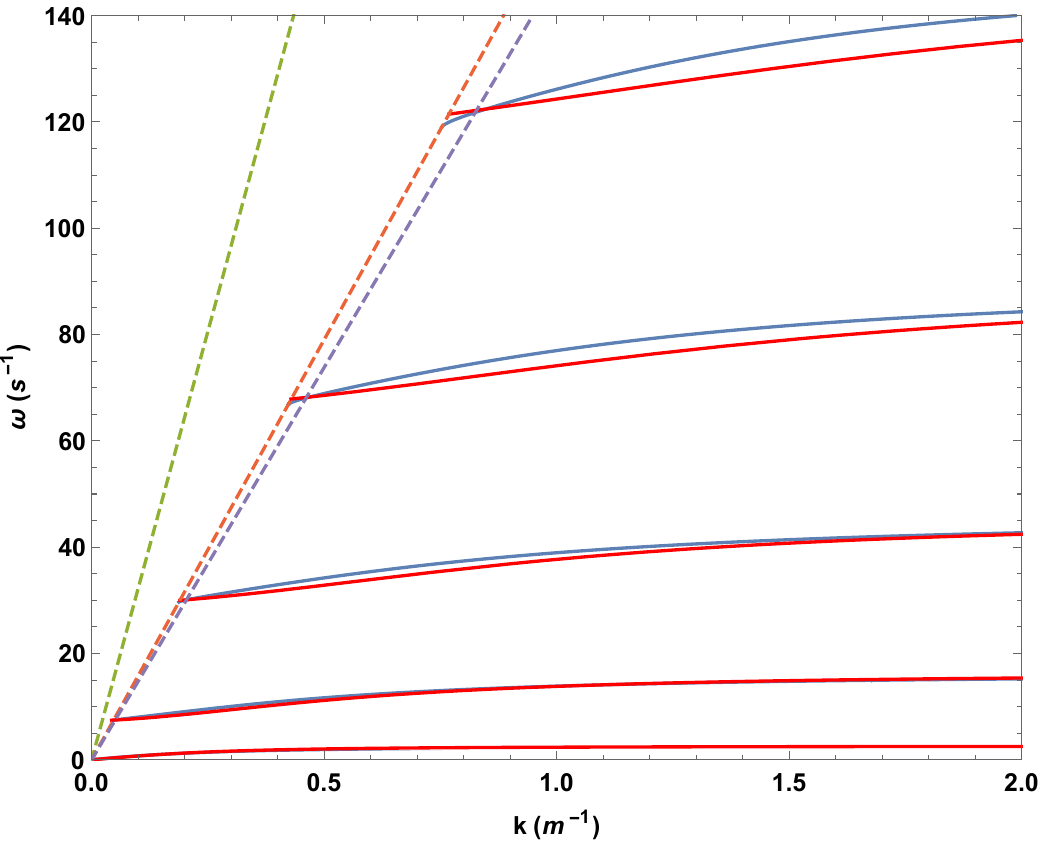}
	};
\draw[very thick,-latex, gray] (-6.5,-1.25) to [bend right] (-3.25,-1.725);
\draw[very thick,-latex, gray] (-5.5,1.5) to [bend left] (-2.05,0.25);
\draw[very thick,-latex, gray] (6,0) to [bend left] (-0.4,3.1);
\node at (-6.5,-1){
	\includegraphics[height=0.5\linewidth]{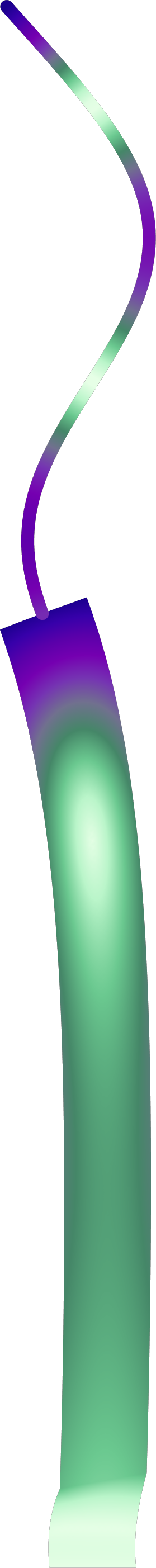}
	};
\node[draw=black, thick, circle, anchor=center] at (-5.75,-4.25){
	\small\sffamily\textbf{a}
	};
\node at (6,0){
	\includegraphics[height=0.5\linewidth]{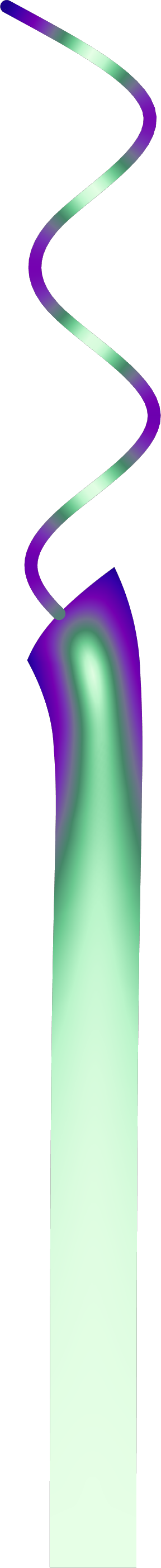}
	};
\node[draw=black, thick, circle, anchor=center] at (5.5,-3){
	\small\sffamily\textbf{c}
	};
\node at (-5.5,2){
	\includegraphics[height=0.5\linewidth]{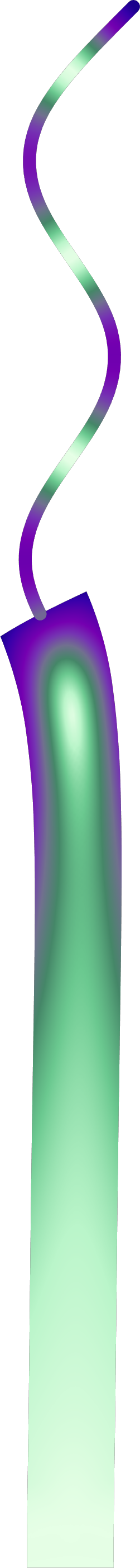}
	};
\node[draw=black, thick, circle, anchor=center] at (-4.9,-0.8){
	\small\sffamily\textbf{b}
	};

\end{tikzpicture}
\caption{The dispersion curves for surface waves on the half-space coated with fully matched beam like resonators, using physical parameters from Table \ref{Tab:BeamsandHalfPlaneParameters}. The solid blue lines show the dispersion curve of the full unimodal solution from \eqref{eq:FullyMatchedExactDispRel} and the solid red lines the dispersion curve of the asymptotic solution with stress interaction term from \eqref{eq:FullyMatchedAsymptoticDispRelWithDet}. The dashed green, orange and purple lines correspond to the longitudinal, shear and Rayleigh wave lines respectively.
Alongside is shown the corresponding mode shapes for each branch; the deformation shows the mode shape and the colour-scheme indicates the magnitude of the total displacement from white (zero) through green to purple (maximum).}
\label{Fig:FullyMatchedBeamResonatorSurfaceWavesWithDet}
\end{figure}
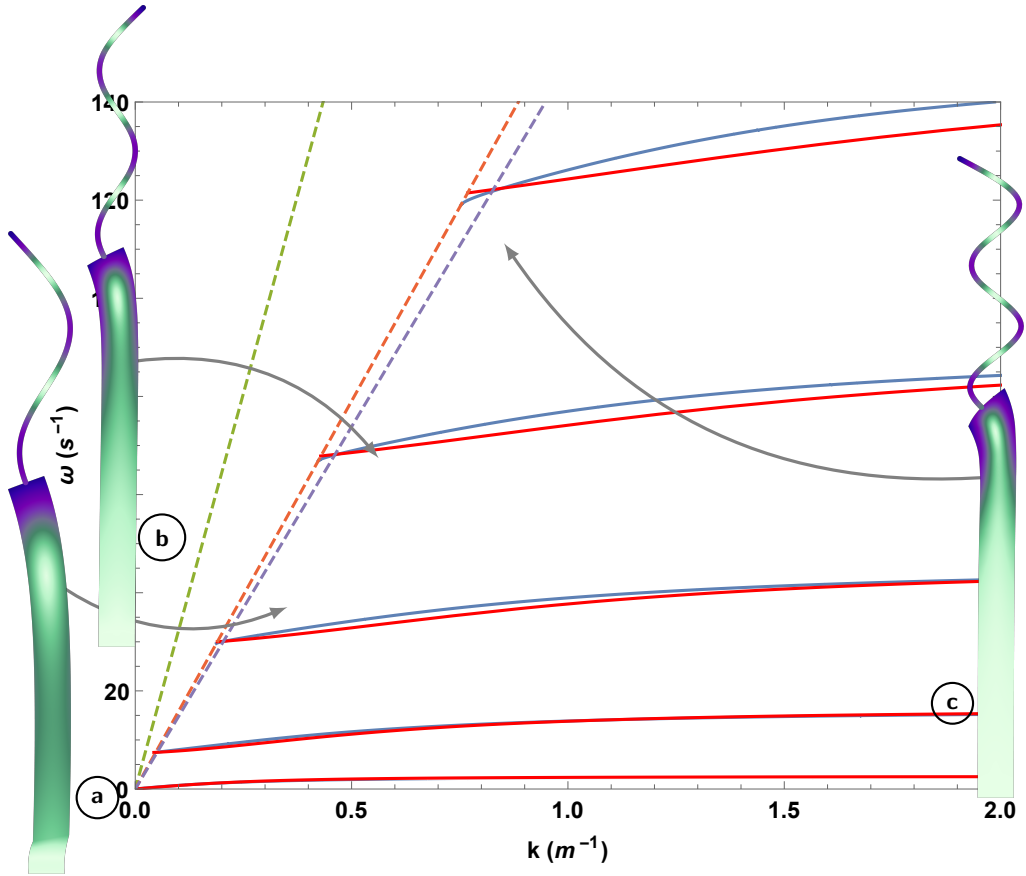

Figure \ref{Fig:FullyMatchedBeamResonatorSurfaceWavesWithDet} shows the dispersion curves associated with the full unimodal solution~\eqref{eq:FullyMatchedExactDispRel} and those arising from the asymptotic approach accounting for the stress interaction~\eqref{eq:FullyMatchedAsymptoticDispRel}; this should be compared with Figure~\ref{Fig:FullyMatchedBeamResonatorSurfaceWaves}.
Figure \ref{Fig:FullyMatchedBeamResonatorSurfaceWavesWithDet} illustrates that accounting for the stress interaction forces the model to coincide with the full unimodal dispersion relation at the Rayleigh line.
That said, comparing Figures~\ref{Fig:FullyMatchedBeamResonatorSurfaceWaves} and \ref{Fig:FullyMatchedBeamResonatorSurfaceWavesWithDet} we observe that the dispersion curves arising from the asymptotic approach where the stress interaction is taken into account
does not show a significant improvement for the rest of the plot, away from the Rayleigh line, over the standard asymptotic model.
This indicates that, for this system, the stress interaction does not have a significant effect on the overall behaviour of the waves away from the Rayleigh line.

Again, comparing Figure~\ref{Fig:SimplySupportedBeamResonatorSurfaceWaves} with Figures~\ref{Fig:FullyMatchedBeamResonatorSurfaceWaves} and~\ref{Fig:FullyMatchedBeamResonatorSurfaceWavesWithDet}, it is apparent that the fully matched junction condition allows for the creation of significantly larger deep sub-wavelength band gaps compared with the simply supported junction condition.
Therefore, the effect of the flexural resonances on the overall behaviour of the system can be increased by an appropriate choice of junction condition.

A subtlety associated with applying the asymptotic model to multi-stress systems is that, the asymptotic approach initial developed in~\cite{kaplunov2004asymptotic} requires that the overall stress is small.
It follows that for multi-stress systems, all individual stresses must be small in for the approach to be applicable.
If one stress is always large while the other is small then the asymptotic model cannot be valid, even if the result crosses the Rayleigh line.

Finally, away from the Rayleigh line, the dispersion curves approach asymptotes associated with the flexural resonances of the beams; this is a known feature of sub-wavelength resonant arrays (see, for example,~\cite{williams2015theory}).
For the previous two cases of simply supported and rail junction conditions, the asymptotic model correctly reproduces the asymptotes.
This led to the asymptotic model remaining accurate to a good degree even away from the Rayleigh line.
In contrast, the fully matched junction (both uni- and multi-stress) does not accurately reproduce the necessary asymptotes.
In particular, the asymptotes of the full unimodel solution are solutions of the transcendental equation
\begin{align}
\cos(KL)\cosh(KL)+1 &= 0.
\end{align}
However, the fully matched asymptotic models produces asymptotes at,
\begin{align}
-\frac{1+\beta_R^2}{2\mu B l^2}\frac{\pi h^4}{64} E_R K^2 \frac{1 - \beta_R^2}{2}\ \sin(KL)\sinh(KL) = \cos(KL)\cosh(KL) +1.
\end{align}
It is clear that, the fully matched asymptotic solutions and the full unimodal solution will diverge as $k$ becomes large.
Therefore, whilst this is a powerful technique for when both stresses are small, the results cannot be relied upon away from this condition.

\begin{figure}
\centering
\begin{subfigure}{0.03\linewidth}
\centering
\includegraphics[width=\linewidth]{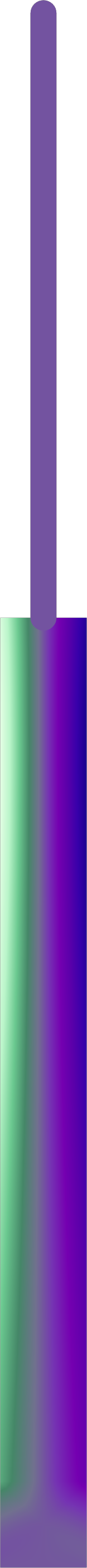}
\caption{}
\end{subfigure}
\qquad\qquad
\begin{subfigure}{0.03\linewidth}
\centering
\includegraphics[width=\linewidth]{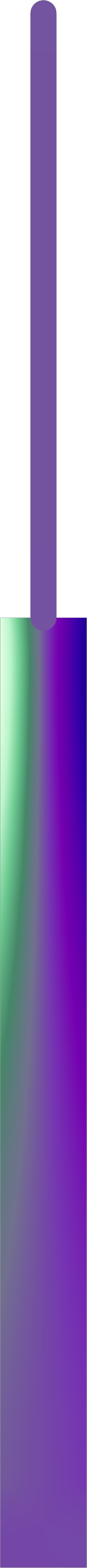}
\caption{}
\end{subfigure}
\qquad\qquad
\begin{subfigure}{0.03\linewidth}
\centering
\includegraphics[width=\linewidth]{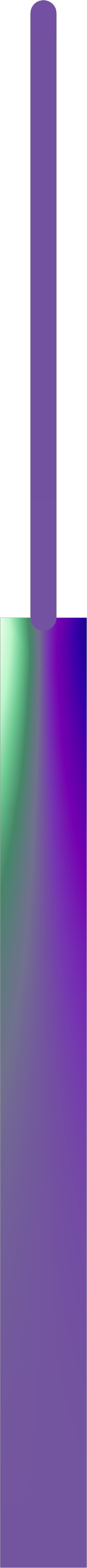}
\caption{}
\end{subfigure}
\caption{The vertical ($u_3$ and $U_3$) components of the displacement field for the three modes shown in Figure~\ref{Fig:FullyMatchedBeamResonatorSurfaceWavesWithDet}.
The colour-scheme indicates the size of the vertical component of the displacement field from white (minimum) through green to purple (maximum).}
\label{Fig:FEM-vertical}
\end{figure}

Figure \ref{Fig:FullyMatchedBeamResonatorSurfaceWavesWithDet} also shows the mode shapes associated with each branch of the dispersion curves; the colour scheme shows the magnitude of the total displacement.
Animated versions of these figures can be found in the supplementary material.
The modes were computed, with the Finite Element software COMSOL Multiphysics\textsuperscript{\textregistered}, using the exact formulation for an infinite periodic array of resonators as described in \S\ref{subsec:An Array of Beams embedded into an Elastic Half Plane}.
It can be observed that each mode can be associated with the fundamental modes of the flexural resonators; this is particular apparent from the animated mode shapes that can be found in the supplementary material.
Indeed, it is clear that the flexural motion of the beams dominates the behaviour of the system in this regime and emphasises that the flexural resonances can play an important part in such problems where slender beams rest atop an elastic half-space.

Moreover, Figure~\ref{Fig:FEM-vertical} shows the vertical components of the displacement in the beam ($U_3$) and half-space ($u_3$).
It is clear that, in terms of compressional motion, the beams move as a rigid body in the vertical direction and, therefore, the dynamic effects of the compressional motion of the beams do not materially contribute to the overall behaviour of the system in this regime.
This observation is particularly evident in the animated versions of Figure~\ref{Fig:FEM-vertical}, which are available in the supplementary material.

\section{Concluding remarks}\label{section:Concluding remarks}

An array of flexural resonators attached to the surface of an elastic half-space is analysed using an explicit model for the Rayleigh wave.
The paper generalises previous considerations using both full unimodal~\cite{colquitt2017seismic} and asymptotic solutions for an elastic half-space in the case of an  array of compressional resonators~\cite{ege2018approximate}.

Three types of junction conditions are considered, and for each the asymptotic solution is verified using full unimodal solutions for the half-space.
The first junction condition to be considered treats the resonators as simply supported at the surface of the half-space.
Matching of the beam equation yields a horizontal stress only.
The resulting dispersion curves exhibit a series of very fine band gaps typical of deep sub-wavelength arrays; the narrow nature of these band gaps means that making use of these stop bands in practical applications such as filtering and mode-conversion is extremely challenging.
For example, deep sub-wavelength arrays cannot be used to design broadband filters.

The second considered junction condition treats the resonators as being supported by a freely moving rail at the surface of the half-space.
In contrast to the previous system this yields a vertical stress only.
The effect of this junction condition is to widen the band gaps in the spectrum.
This ability to increase the width of the stop bands through a judicious choice of the junction conditions is an important feature that would allow, for example the implementation of broadband filters in the deep sub-wavelength regime.
We also note that this effect will increase the contribution of the dynamic flexural behaviour of the resonators on the overall response of the system.
Whilst sub-wavelength resonant arrays have found extensive use in the literature (see, for example,~\cite{zhou2012elastic}), little attention has thus far been given to the junction condition joining the resonators to the substrate.
Indeed, one application of the present paper is in the analysis of the interaction of surfaces waves with dense pine forests and recent experimental data suggests that the flexural resonances as well as the junction conditions may be an important factor~\cite{roux2018toward}.

For both the simply supported and rail junction conditions the asymptotic formulation shows good agreement with the full unimodal solution.
The asymptotic solution obtained predicts both a coincidence with the full unimodal solution at the Rayleigh line, and also resonances arising from the solution to the beam equation.
Moreover, the asymptotic solution has the distinct advantage of being much simpler and easier to manipulate than the full unimodal solution.

The final junction condition considered in this paper fully matches all variables at the junction.
Unlike the previous two cases this results in both a vertical stress and a horizontal stress, requiring an alteration to the existing asymptotic formulation to account for both stresses
While this asymptotic solution closely matches with the full unimodal solution it does not accurately predict the coincidence with the Rayleigh line or the location of the system resonances.
A further addition to the asymptotic model, which accounts for the coupling of stresses at the junction point, produces a solution which does accurately predict coincidence with the full unimodal solution with the Rayleigh line, but still does not accurately predict resonances.

All of the systems considered produce band gaps in the region around the beam resonances, with the largest band gaps produced by the beam on a rail and fully matched junction conditions.

This is an important feature as, usually, sub-wavelength resonant arrays are only capable of producing very narrow band gaps~\cite{zhou2012elastic} and, as a result, it is challenging to make use of these band gaps in the design of filters and other control devices.
However, as we have demonstrated, it is possible to control the width of these deep sub-wavelength band gaps by adjusting the junction condition joining the resonators to the half-space.
Indeed, the beam on a rail and full matched junction conditions lend themselves to feasible designs of mechanical structure that exhibit large band gaps and are capable of controlling the propagation of surface waves.
In particular, the beam on a rail junction condition is associated with gyroscopic resonators, as discussed in~\cite{nieves2018vibrations}, and which can be designed to exhibit the desired properties;
the fully match condition is, in some sense, the most natural junction condition to impose as it most closely resembles the conditions of ideal contact.

This demonstrates the potential for such systems to be utilised in controlling the propagation of surface waves. 

The explicit asymptotic formulation has the potential to be applied in more sophisticated systems where a full unimodal solution may be difficult or impossible to obtain since it reduces the vector problem in full linear elasticity to a scalar problem along the surface.
It has been shown that for the systems considered the asymptotic formulation easily produces simple explicit solutions which match closely with the full unimodal solution. 

The formulation discussed involves a surface scalar problem identical to the equation for transverse forces applied to an elastic membrane.
Currently, experimental set-ups to model elastic half-spaces can be large and expensive~\cite{colombi2017elastic} and this formulation gives scope for simpler experiments on membranes which model the behaviour at the surface of an elastic half-space.

This system could also be further developed by considering the near-surface effect of coupled stresses. We also remark that the effect of microstructure, including non-local effects, have previously been considered for near-surface wave motion \cite{chebakov2016refined} and, as might be expected,  are asymptotically secondary for treating Rayleigh-type waves.

\subsection*{Funding}
D. J. C. gratefully acknowledges the financial support of the EPSRC (UK) through programme grant no. EP/L024926/1. P.T.W also acknowledges Keele University for supporting their PhD studies.

\bibliographystyle{ieeetr}
\bibliography{surface-waves-refs}

\end{document}